\documentclass[twocolumn,reprint, secnumarabic,amssymb, nofootinbib, notitlepage, aps, prd,longbibliography]{revtex4-1}

\usepackage{amssymb,amsmath,amsfonts,mathrsfs,graphicx,xcolor,units,yfonts}
\usepackage{hyperref}

\hypersetup{
  colorlinks=true,        % false: boxed links; true: colored links
  linkcolor=blue,         % color of internal links
  citecolor=blue,       % color of links to bibliography
  urlcolor=black  }
\newcommand{\dmunu}{_{\mu\nu}}
\newcommand{\umunu}{^{\mu\nu}}
\newcommand{\enr}{\mathcal{E}}
\newcommand{\prs}{\mathcal{P}}

\begin{document}

%\preprint{}

\title{Master equation as a radial constraint}


\author{Uzair Hussain}
\email[]{uh1681@mun.ca}
\author{Ivan Booth}
\email[]{ibooth@mun.ca}
\author{Hari K. Kunduri}
\email{hkkunduri@mun.ca}
%\thanks{}
\affiliation{Department of Mathematics and Statistics,   Memorial University of Newfoundland  St John's NL A1C 4P5, Canada}


\date{\today}

\begin{abstract}
We revisit the problem of perturbations of Schwarzschild-AdS$_4$ black holes by using
 a combination of the Martel-Poisson formalism for perturbations of four-dimensional spherically symmetric spacetimes \citep{Martel:2005ir} and the Kodama-Ishibashi
 formalism  \citep{Kodama:2003jz}. We clarify the relationship between both formalisms and express the Brown-York-Balasubramanian-Krauss boundary stress-energy tensor, $\bar{T}_{\mu\nu}$, 
 on a finite-$r$ surface purely in terms of the even and odd master functions. Then, on these surfaces we find that the spacelike components of the conservation equation 
  $\bar{\mathcal{D}}^\mu \bar{T}_{\mu\nu} =0$ are equivalent to the wave equations for the master functions. 
%     % along the {\color{green} directions tangent to spherical symmetry}. % This direct comparison of the master equations and the conservation equations is presented explicitly for the first time. 
{ The renormalized stress-energy tensor at the boundary $\displaystyle \frac{r}{L} \lim_{r \rightarrow \infty} \bar{T}_{\mu\nu}$ is calculated directly in terms of the master functions}.
% using our expressions for the stress-energy tensor. %and demonstrate the fluid/gravity duality for large black holes.
 % We also investigate the possibility of a Cotton tensor/stress-energy tensor duality, on a finite-$r$ surface, motivated by the duality being held at $r\rightarrow\infty$ [arXiv:0809.4852v2].
\end{abstract}

\maketitle
\section{Introduction}
The linear perturbation theory of %asymptotically flat 
Schwarzschild spacetimes has been applied to a wide range of physical scenarios such as the prediction of gravitational radiation, stability analysis, studying binary systems, and the scattering and absorption of gravitational radiation \cite{Frolov:1998wf}. 
Since its inception by Regge and Wheeler \cite{Regge:1957td} as a tool for studying the stability of Schwarzschild black holes, 
the perturbation formalism has received steady enhancements. Early fundamental contributions were made by Zerilli \cite{Zerilli:1971wd}, Vishweshwara \cite{Vishveshwara:1970cc} and 
Chandrasekhar \cite{Chandrasekhar:1985kt}. Although powerful, the equations were limited to particular gauge choices under infinitesimal coordinate transformations: the well-known Regge-Wheeler, 
and Zerilli gauges. This lack of gauge invariance was remedied by Moncrief in  \cite{Moncrief:1974am} where the equations were presented in a gauge invariant formalism. Further upgrades to a 
coordinate independent formalism were made by Gerlach and Sengupta \cite{Gerlach:1980tx}. 

%More recently all of this work has been unified  in an {\color{red} independent}
% pair of  {\color{red} gauge invariant and coordinate independent} formalisms. 
{ More recently, there have been two further generalizations which incoporate gauge invariance and coordinate independence.}
Martel and Poisson developed a particularly robust and 
practical four-dimensional formalism in \cite{Martel:2005ir} which also included the linear effect of matter sources. Meanwhile, in \cite{Kodama:2003jz} Kodama and Ishibashi
generalized to perturbations of any maximally symmetric black hole in spacetime dimensions $d \ge 4$.
% 
%Given the importance of the theoretical laboratory provided by the AdS/CFT correspondence \cite{Maldacena:1997re} there has also been tremendous interest in studying perturbations of asymptotically AdS spacetimes. Of particular interest are the quasi-normal modes (QNMs) of such black holes. The QNMs for a scalar field around an AdS-Schawarschild black hole were first calculated by Horowitz and Hubeny \cite{Horowitz:1999jd} by using a power series method. QNMs for a scalar field in Reissner-Nordstrom AdS black holes were calculated by Wang et. al. \cite{Wang:2000gsa} by using the method of \cite{Horowitz:1999jd}. A complete study of the QNMs of gravitational and electromagnetic perturbations was obtained by Cardoso and Lemos \cite{Cardoso:2001bb}. Since the goal was the QNMs, these calculations were all done in the frequency domain %and in a particular 
%with a fixed gauge. A superb gauge invariant and coordinate independent formalism which generalizes perturbations of all maximally symmetric black holes in spacetime dimension $d\ge 4$, was achieved by Kodama and Ishibashi \cite{Kodama:2003jz}. 

In the current work we apply the formalisms developed in these two papers to study aspects of the AdS/CFT correspondence. We are especially interested in the body of work flowing from 
the calculation of the effective shear viscosity of the gauge theory in the strongly coupled regime at finite temperature \cite{Policastro:2001yc}. The marrow of that calculation was the observation that an interacting quantum field theory under local thermal equilibrium can be effectively described in terms of fluid dynamics\cite{Bhattacharyya:2008jc}. In this regime the AdS/CFT correspondence can be viewed as a fluid/gravity correspondence by looking at long wavelength fluctuations about equilibrium (see \cite{Hubeny:2011hd} and \cite{Rangamani:2009xk} and references therein). 

In this regime the fields on both sides of the duality are classical and so it can be established independently without recourse to more general arguments. Directly from general relativity, one 
may identify the Brown-York-Balasubramanian-Krauss  (BYBK) stress-energy tensor \cite{Balasubramanian:1999re} induced at timelike infinity with the stress-energy tensor of a 
near-ideal fluid. In such a setting one may compare the perturbations of black holes/branes with corresponding perturbations of the fluid velocity, energy, and pressure. 

For five-dimensional AdS$_5$ black-brane spacetimes, a systematic procedure to study this correspondence was developed by Bhattacharyya et. al. \cite{Bhattacharyya:2008jc}.
The approach begins by writing an equilibrium brane solution coordinate-boosted to the proper velocity of the boundary fluid. One then perturbatively solves the Einstein
equations order-by-order over the background metric in terms of derivatives of the boundary fluid velocity and temperature. In analogy to the (3+1) formulation of general relativity, the 
 Einstein equations can be decomposed into constraints on (timelike) constant coordinate-radius surfaces along with radial evolution equations.
%
%be viewed as radial evolution equations and radial constraint equations, which follow from the Gauss-Codacci relations. 

Now, even away from infinity, one can calculate a quasilocal BYBK stress-energy tensor on each surface of constant coordinate-radius.
A crux of the calculation is that the diffeomorphism-constraint equation on each constant-radius surface is identical with the conservation of the induced stress-energy tensor 
along that surface\footnote{The conservation law follows directly from the Gauss-Codacci equations. From the geometric perspective it is an identity which holds on any timelike surface. }. 
Meanwhile the radial evolution equation ensures that such surfaces link together to form a coherent spacetime.  

In \cite{Bhattacharyya:2008jc} this formalism was worked out for AdS$_5$ black branes up to second order in derivative expansion. Since the behaviour of 2+1 dimensional fluids is different, especially in terms of the behaviour of turbulence, Raamsdonk in \cite{VanRaamsdonk:2008fp}, applied the same methods to AdS$_4$ black branes again up to second order in derivatives.

%An exciting question that arises in such a relationship between fluids and gravity is; if fluids are turbulent does that mean gravity is turbulent also? One exploration of this idea comes from \cite{Yang:2014tla}, where it was shown that asymptotically flat Kerr black holes in space time dimension $d=4$, display turbulence that is driven by a parametric instability which transfers energy from short wavelengths to long wavelengths \footnote{There is also strong evidence that pure AdS is nonlinearly unstable \cite{Bizon:2013gxa}}. Such an inverse cascade is also observed for asymptotically AdS$_4$ spacetimes \cite{Adams:2013vsa}. Inverse cascades are a property of fluids in 2+1 dimensions \cite{Carrasco:2012nf} and thus this behaviour in the bulk makes sense in terms of the fluid/gravity duality. Another interesting study concerning turbulence is \cite{Green:2013zba}, where a numerical, turbulent fluid solution was mapped into the bulk and the same inverse cascades were found. A geometric realization of the Reynolds number in the bulk was also sought for \cite{Green:2013zba}. 

In this current paper we will be concerned with how the fluid/gravity duality arises for large\footnote{Recall that only black holes whose mass is large relative to the radius of 
cosmological curvature are thermodynamically stable\cite{Hawking:1982dh}. Black branes are all large but spherical Schwarzschild-AdS black holes can be either large or small. } spherical AdS$_4$ black holes. Some work has already been done in this area. For example a
 connection between the bulk dynamics of the spherical black hole and the boundary fluid has been made in terms of the quasinormal modes (QNMs) of the black hole. In 
 \cite{Michalogiorgakis:2006jc} the QNMs of the perturbations expanded in even spherical harmonics were computed  using a Robin boundary condition. The authors showed that there were low lying modes which, for 
 large black holes, corresponded precisely to the modes of a linearly perturbed fluid on $\mathbb{R}\times S^2$, the boundary manifold under said Robin boundary conditions. For general 
 boundary conditions the fluid/gravity duality in terms of the boundary BYBK stress-energy tensor is presented in \cite{Bakas:2008gz} (and  further considered in \cite{AidaMSc}) 
 for both even and odd spherical harmonics.

%{\color{red} Can we say something about large black holes here? ie why large, how it shows up in our equations}

Our goal is to understand how the well-developed perturbation theory of spherical black holes in AdS$_4$ is connected to the dynamics of the fluid. In particular we are interested
in understanding the role of the master function on the fluid dynamics side: one of the most remarkable features of the perturbative formalism  is that allowed perturbations of the spacetime are 
determined by a scalar master function which  obeys an inhomogeneous wave equation \cite{Martel:2005ir,Kodama:2003jz}. The whole system of Einstein's equations can be  
characterized by this master variable along with equations that relate it back to the components of the metric perturbation. 

We will show that this master equation is equivalent to the conservation of the quasilocal BYBK stress-energy tensor on finite-$r$ surfaces. This can be thought of as a (non-trivial)
extension of the result from the black-brane formalism\citep{Bhattacharyya:2008jc}, where the radial constraint equation was shown to be equivalent to the conservation equation of the 
induced stress-energy tensor 
and the rest were radial evolution equations. Here, in the spherical case, we show that if we rewrite the metric perturbations in term of the master function then the 
conservation of the induced stress-energy  is equivalent to the master equation.  This can be contrasted to the work of \cite{Brattan:2011my}, in which it was shown that prescribing a Lorenzian metric on a constant-$r$ surface could be used to determine the bulk black brane spacetime metric in the long wavelength regime.

We also show how the form of the BYBK stress-energy tensor is greatly simplified when expressed in terms of the master variable. We provide formulas for the energy, pressure, velocity, viscosity, and vorticity in terms of the master variable both in the bulk and at the boundary. This enables us to express the quantities in the time domain rather than the frequency domain. Lastly, we go to the frequency domain to demonstrate how the fluid at the boundary arises for large black holes.

The paper is organized as follows. Section II reviews standard perturbation theory for spherical black holes in AdS$_4$. Section III considers the stress-energy tensor induced on finite-$r$ surfaces and shows that the conservation equations are equivalent to the master equations derived in the previous section.  Section IV shows how properties of the fluid (e.g. energy, pressure) can be identified in terms of the master function.  We discuss some open problems in Section V.

\section{Perturbations of AdS$_4$ Black holes}
The Schwarzschild AdS$_{4}$ black hole is a solution to Einstein's
equations with a negative cosmological constant $\Lambda < 0$,
\begin{align}
R_{\alpha\beta}= & \Lambda g_{\alpha\beta}
\end{align}
and the metric exterior to the event horizon is given by
\begin{align}
ds^{2}= & -f(r)dt^{2}+\frac{1}{f(r)}dr^{2}+r^{2}(d\theta^{2}+\text{sin}^{2}\theta d\phi)
\end{align}
with, $f(r)=  1-\frac{2M}{r}+\frac{r^{2}}{L^{2}}$ and $L=\sqrt{\frac{-3}{\Lambda}}$. The metric is stationary and spherically symmetric, with $-\infty < t < \infty$, $0 < \theta < \pi$, $0 < \phi < 2\pi$ and $r > r_+$ where $r_+$ is the largest root of $f(r)$. The spacetime is asymptotically AdS$_{4}$ with length scale $L$.
Following \cite{Martel:2005ir}, and given the spherical symmetry
of the spacetime, the metric is expressed as,
\begin{align}
^4g_{\alpha \beta} dX^\alpha dX^\beta= & g_{ab}dx^{a}dx^{b}+r^{2}\Omega_{AB}d\theta^{A}d\theta^{B}.
\end{align}
Here, $^4g_{\alpha \beta}$ is the full 4-dimensional metric, $g_{ab}$ is the metric on the 2-dimensional submanifold $\mathcal{M}^{2}$, consisting of the orbits of spherical symmetry
which in Schwarzschild coordinates is the spatio-temporal or `$(r,t)$'
part. Lastly, $\Omega_{AB}$ are the components of the metric of a unit sphere, $S^{2}$. The 4-dimensional coordinates are expressed as $X^{\alpha}$, the coordinates on $\mathcal{M}^{2}$ are expressed as $x^{a}$ and the coordinates on the sphere are expressed as $\theta^{A}$.
Note that, $\{\alpha,\beta\}$ run over all coordinates, lower-case Latin indices run over $r$ and $t$, and upper case
Latin indices run over $\theta$ and $\phi$. The covariant derivative compatible with $g_{ab}$ will be written as $\nabla_a$ and the covariant derivative compatible with $\Omega_{AB}$ will be written as $D_A$.
Again following \cite{Martel:2005ir} we will introduce the one-form $r_{a}$, 
\begin{align}
r_{a}= & \frac{\partial r}{\partial x^{a}}
\end{align}
which is $r_{a}=(0,1)$ in Schwarzschild coordinates.

\subsection{Odd Perturbations}
We may now perturb the black hole given by adding a perturbation, $p_{\alpha \beta}$. We shall expand this perturbation in terms of odd spherical harmonics, $X_A$ and $X_{AB}$. Their precise definition  can be found the Appendix. In what follows we closely follow \cite{Martel:2005ir} until (\ref{rwcmf}). The perturbation is written as $^4 g_{AB} = r^2 \Omega_{AB} + p_{AB}$ and $^4 g_{aB} = p_{aB}$, where,
\begin{align}
p_{aB}=  \sum_{lm}h_{a}^{lm}X_{B}^{lm}\;, \qquad  p_{AB}=  \sum_{lm}h_{2}^{lm}X_{AB}^{lm}
\end{align}
where $h_a$ and $h_2$ are functions of $x^a$. Infinitesimal gauge transformations will also be expanded in terms of odd harmonics,
\begin{align}
e_A=\sum_{lm} e^{lm} X_A^{lm}
\end{align}
with $e^{lm}$ as a function of $x^a$. Under such gauge transformations we have, dropping the $lm$ indices, the following gauge invariant variables,
\begin{align}\label{oddGI}
\tilde{h}_{a}= & h_{a}-\frac{1}{2}\nabla_{a}h_{2}+\frac{1}{r}r_{a}h_{2}.
\end{align}
All gauge invariant quantities will have the `$^{\sim}$' symbol hereafter. Using the linearized Einstein's equations we find that the whole system is characterized by the following equation,
\begin{equation}\label{masterodd}
(\square -V_{odd})\tilde{\Xi}_{RW}=0
\end{equation}
where $\square$ is the d' Alembertian on $\mathcal{M}^2$, $\tilde{\Xi}_{RW}$ is the well known Regge-Wheeler master function and,
\begin{equation}
V_{odd}=\frac{\lambda}{r^2}-\frac{6M}{r^3}
\end{equation}
for $\lambda=l(l+1)$. Further, in Schwarzschild coordinates, one may reconstruct the metric perturbations from the following equations,
\begin{align}
\tilde{h}_{t}= & f \int \partial_r\left(r\tilde{\Xi}_{RW}\right)dt' \label{psirwdef1} \ \  \text{and} \\
\tilde{h}_{r}= & \frac{r}{f}\tilde{\Xi}_{RW}. \label{psirwdef2}
\end{align}
Note that this system is undetermined and so one needs to pick a gauge to fully reconstruct the perturbation. We will work in the Regge-Wheeler gauge with $h_2=0$.

One can also define an alternate master variable, the Cunningham-Moncrief-Price function $\tilde{\Psi}$,
\begin{equation}\label{CMF}
\frac{\mu}{2r}\tilde{\Psi}=  \left(\partial_{r}\tilde{h}_{t}-\partial_{t}\tilde{h}_{r}-\frac{2}{r}\tilde{h}_{t}\right)
\end{equation}
where $\mu = (l-1)(l+2)$. This is related to $\Xi_{RW}$ by
\begin{equation}\label{rwcmf}
\tilde{\Xi}_{RW}= \frac{1}{2}\partial_{t}\tilde{\Psi}.
\end{equation}
Interestingly, $\tilde{\Psi}$ satisfies the same master equation, (\ref{masterodd}), as $\tilde{\Xi}_{RW}$.

We can compare the above results with those from the formalism of \cite{Kodama:2003jz} by noting the following relationships between their notation and the one used here. Comparing the metric perturbations we find
\begin{equation}
h_a \leftrightarrow -rf_a \qquad \text{and}\qquad
h_2 \leftrightarrow \frac{2r^2}{k_V}H_T
\end{equation}
which leads to the following relationship for the gauge invariant variables
\begin{equation}
\tilde{h}_a \leftrightarrow -rF_a.
\end{equation}
The master function in \cite{Kodama:2003jz} is defined by
\begin{equation}\label{ishodd}
rF^a=\epsilon^{ab}\partial_b(r\tilde{\Psi}_{KI}).
\end{equation}
Comparing (\ref{ishodd}) with (\ref{psirwdef1}) and (\ref{psirwdef2})  it can be deduced that
\begin{equation}
-2\tilde{\Psi}_{KI}=\tilde{\Psi}.
\end{equation}
So the master function used in \cite{Kodama:2003jz} is essentially the same as that of CMF.

\subsection{Even Perturbations}
Following the same scheme as for the odd perturbations and \cite{Martel:2005ir}, we write the perturbation $p_{\alpha \beta}$ as,  $^4 g_{ab} = g_{ab} + p_{ab}$, $^4 g_{AB} = r^2 \Omega_{AB} + p_{AB}$ and $^4 g_{aB} = p_{aB}$. Now the perturbations will be expanded in even harmonics, $Y^{lm}$, $Y_A^{lm}$, $Y_{AB}^{lm}$, and $\Omega_{AB} Y^{lm}$. The definitions of these can be found in the Appendix. Then the perturbations are
\begin{align}
p_{ab}= & \sum_{lm}h_{ab}^{lm}Y^{lm},\\
p_{aB}= & \sum_{lm}j_{a}^{lm}Y_{B}^{lm} \ \ \text{and}\\
p_{AB}= & r^{2}\sum_{lm}\left(K^{lm}\Omega_{AB}Y^{lm}+G{}^{lm}Y_{AB}^{lm}\right)
\end{align}
where $h_{ab}^{lm}$, $j_a^{lm}$, $K^{lm}$ and $G^{lm}$ are functions of $x^a$. Infinitesimal gauge transformations are expanded in terms of the even harmonics
\begin{equation}
e_a = \sum_{lm} e_a^{lm} Y^{lm} \qquad \text{and} \qquad
e_A = \sum_{lm} e^{lm} Y_A^{lm} 
\end{equation} 
with $e_a^{lm}$ and $e^{lm}$ as functions of $x^a$. Under such gauge transformations we have, dropping the $lm$ indices, the following gauge invariant variables,
\begin{align}
\tilde{h}_{ab}:= & h_{ab}-\nabla_{a}\varepsilon_{b}-\nabla_{b}\varepsilon_{a}\\
\tilde{K}:= & K+\frac{1}{2}\lambda G-\frac{2}{r}r^{a}\varepsilon_{a}
\end{align}
for,
\begin{align}
\varepsilon_{a}:= & j_{a}-\frac{1}{2}r^{2}\nabla_{a}G.
\end{align}

%{\color{red} It is now convenient to proceed using the master function and formalism from \cite{Kodama:2003jz}} {\color{red}Comparing with \cite{Martel:2005ir}:} 

{ We now proceed using the master function from \cite{Kodama:2003jz}, since in \cite{Martel:2005ir} the treatment is for asymptotically flat  rather than asymptotically AdS black holes. We can make this switch by noting how the notation of the two compare: }
\begin{align}
& h_{ab} \leftrightarrow f_{ab}\; , \quad
j_a \leftrightarrow  -\frac{1}{k} r f_a \;, \quad \\
& K \leftrightarrow 2H_L\;  \quad \text{and} \quad
G \leftrightarrow \frac{2}{k^2} H_T
\end{align}
which leads to relationships for the gauge invariant variables,
\begin{equation}
\varepsilon_a \leftrightarrow -X_a \; , \quad
\tilde{h}_{ab} \leftrightarrow F_{ab} \; \quad \text{and} \quad
\tilde{K} \leftrightarrow2F.
\end{equation}
Then in terms of the functions $X$,$Y$, and $Z$ from \cite{Kodama:2003jz} we have,
\begin{align}
\tilde{h}_{tt}&=-\frac{f}{2} \left(X-Y\right)\label{htt} \; , \quad 
\tilde{h}_{rr}=-\frac{1}{2f} \left(X-Y\right),\\
\tilde{h}_{rt}&=\frac{1}{f}Z \;  \quad \text{and} \quad
\tilde{K}=-\frac{X+Y}{2}.
\end{align} 
The master function is defined by the following equations:
\begin{align}
 X&=\frac{1}{r}\left(
   -\frac{r^2}{f}\partial_t^2\tilde{\Phi} 
   -\frac{ P_X}{16H^2}\tilde{\Phi}
   +\frac{Q_X}{4H}r\partial_r\tilde{\Phi}\right)
   \label{XbyPhi:tr},\\
Y&=\frac{1}{r}\left(
   \frac{r^2}{f}\partial_t^2\tilde{\Phi} 
   -\frac{ P_Y}{16H^2}\tilde{\Phi}
   +\frac{Q_Y}{4H}r\partial_r\tilde{\Phi}\right)
   \label{YbyPhi:tr}\ \ \text{and}\\
Z&=-\frac{P_Z}{4H}\partial_t\tilde{\Phi} 
     +f r\partial_r\partial_t\tilde{\Phi}, \label{Z}
\end{align}
where 
\begin{equation}
H:=\mu+\frac{6M}{r}
\end{equation}
 and $P_X,P_Y,Q_X,Q_Y$ and $P_Z$ are all functions of $r$ as defined in \cite{Kodama:2003jz}. The master function satisfies the following wave equation:
\begin{equation}\label{mastereven}
(\square -V_{even})\tilde{\Phi}=0
\end{equation}
where,
\begin{align}
V_{even} = \frac{1}{H^2} \biggl[&  \mu^2 \biggl(
  \frac{\mu+2}{r^2} + \frac{6M}{r^3} \biggr) 
+ \frac{36M^2}{r^4} \biggl(\mu + \frac{2M}{r} \biggr) \nonumber\\
& +72 \frac{M^2}{r^2L^2} \biggr] .
\end{align}

\section{Stress-Energy Tensor}
In this section we will show how the conservation of the induced quasilocal stress-energy tensor on a finite-$r$ surface is equivalent to the master equation, for both the odd and even perturbations. The formula for the stress-energy tensor is as in the usual Brown-York \cite{Brown:1992br} treatment with Balasubramanian-Krauss counterterms  added to regulate the $r \rightarrow \infty$ divergences for AdS \cite{Balasubramanian:1999re}:
\begin{equation}\label{stresstensor}
\bar{T}\dmunu :=\kappa^{-2} \bar{\mathcal{T}}_{\mu\nu} = \bar{K}_{\mu\nu} - \bar{K} \bar{\gamma}_{\mu\nu} -2 \sqrt{-{\Lambda \over 3}}
\bar{\gamma}_{\mu\nu} + \sqrt{-{3 \over \Lambda}}\ ^{3}\bar{G}_{\mu\nu},
\end{equation}
where the indices $\{\mu,\nu\}$ run over all coordinates but $r$, $\kappa^{-2}$ is a constant, $\bar{\gamma}_{\mu\nu}$ is the metric on the finite-$r$ 3-surface, $\bar{K}_{\mu\nu}$ is the extrinsic curvature and $\bar{K} = \bar{\gamma}\umunu \bar{K}\dmunu$ is its trace, and $^{3}\bar{G}$ is the Einstein tensor for $\bar{\gamma}\dmunu$. The bar notation is there to remind us that the quantity includes a perturbation,  e.g, $\bar{A}_{\mu \nu} = A\dmunu + \delta A\dmunu$.
%\begin{equation}
%G^{3} _{\mu\nu}=R^{3} _{\mu\nu}- \frac{1}{2} R^{3} \gamma\dmunu
%\end{equation}

\subsection{Stress-energy tensor for static AdS$_4$ black holes}
In this section we calculate the stress-energy tensor for the static black hole, i.e., without perturbations. We use Schwarzschild coordinates with the normal vector, 
\begin{equation}
n_\alpha=\frac{1}{\sqrt{f}}\delta^r_\alpha.
\end{equation}
The metric on the timelike slice has components $\gamma_{tt}=-f(r), \gamma_{AB}=r^2 \Omega_{AB}$ Using the following formula for the extrinsic curvature,
\begin{equation}
K_{\alpha \beta}=-\nabla_\alpha n_\beta - n_\alpha n^\gamma \nabla _\gamma n_\beta
\end{equation} 
the non-vanishing components are,
\begin{equation}
K_{tt}=\frac{\sqrt f f'}{2} \;  \quad \text{and} \quad
K_{AB}=-\sqrt{f} r\Omega_{AB}
\end{equation} with trace
\begin{equation}
K=-\frac{f'}{2\sqrt{f}}-\frac{2\sqrt{f}}{r}.
\end{equation}
After including the counter terms shown in (\ref{stresstensor}) the non-vanishing components of the stress-energy tensor are
\begin{align}
&T_{tt}=\frac{1}{r^2}\left( L+\frac{2r^2}{L} - 2r\sqrt{f} \right) f  = \tau_1 f \label{oddstess1} \ \ \text{and}\\ 
&T_{AB}= \left(\frac{f'}{2\sqrt{f}}+\frac{\sqrt{f}}{r} -\frac{2}{L}\right)  r^2\Omega_{AB} = \tau_2 r^2 \Omega_{AB} \label{oddstress2}
\end{align} which defines the functions $\tau_1,\tau_2$. 
%where,
%\begin{align}
%&\tau_1:=\frac{1}{r^2}\left( L+\frac{2r^2}{L} - 2r\sqrt{f} \right)\\
%&\tau_2:=\frac{f'}{2\sqrt{f}}+\frac{\sqrt{f}}{r} -\frac{2}{L}
%\end{align}

\subsection{Conservation of odd stress-energy tensor}
In this section we will calculate the the odd perturbation of the stress-energy tensor and demonstrate the equality of the conservation equation and the odd master equation (\ref{masterodd}). To calculate the perturbation  note that since we are taking traces with $\bar{\gamma}\umunu= \gamma \umunu - \delta \gamma \umunu $, the trace of an unperturbed quantity will pick up a perturbation, for e.g., $\bar{A}=\gamma \umunu A \dmunu - \delta \gamma  \umunu A \dmunu$.  The expression for the odd perturbation to the stress-energy tensor in a general gauge is,
\begin{align}
\delta T_{tA}= & \frac{\sqrt{f}}{2}\left(\partial{}_{t}\tilde{h}{}_{r}-\partial_{r}\tilde{h}_{t}+\frac{2}{r}\tilde{h}_{t}\right)X_{A} \\
& +\frac{L\mu}{2r^{2}}\tilde{h}_{t}X_{A}+\tau_2h_{t}X_{A} \ \ \text{and} \nonumber \\
\delta T_{AB}= & \left(\sqrt{f}\tilde{h}_{r}-\frac{L}{f}\partial_{t}\tilde{h}_{t}\right)X_{AB}+\tau_2h_{2}X_{AB}.
\end{align}
{ These terms cannot be written purely in terms of the gauge independent $\tilde{h}_a$ (\ref{oddGI}) and so the quasilocal stress-energy is gauge dependent. 
However, this dependence does not effect the conservation equations: they hold for all gauges.}
%whatever the gauge the conservation equations  (\ref{conodd}) hold.}% and so we write them with a `$^{\sim}$' on top. }

%Working in a general gauge we were able to collect together terms that {\color{red} contain} %form 
%the gauge invariant quantity (\ref{oddGI}) and so find that the full expressions are not gauge invariant.

This invariance allows us to freely choose a gauge. We choose the Regge Wheeler gauge $h_{2}=0$, $\tilde{h}_{t}=h_{t}$ and $\tilde{h}_{r}=h_{r}$ and use (\ref{psirwdef1}) and (\ref{psirwdef2}) to express the stress-energy tensor in terms of $\Xi_{RW}$. Now, we invoke the conservation equations
\begin{equation}\label{conodd}
\mathcal{\tilde{Q}}_\nu:=\bar{\mathcal{D}}^\mu \bar{T}_{\mu\nu} = 0 .
\end{equation}
Here the $\bar{\mathcal{D}}$ is the covariant derivative compatible with $\bar{\gamma}\dmunu$, the bar on $\mathcal{D}$ indicates that the Christoffel symbol contains a perturbation. The index is raised with metric plus its perturbation, $\tilde{\gamma}_{\mu\nu} = \gamma_{\mu\nu} + \delta\gamma_{\mu\nu}$.
Keeping only the linear terms we find that the $\nu=t$ equation of (\ref{conodd}) is trivially satisfied, whereas the $\nu=A$ equations result in
\begin{equation}
\left(\square  - V_{odd}  \right)\tilde{\Xi}_{RW} X_A = 0
\end{equation}
which is equivalent to the Regge-Wheeler master equation (\ref{masterodd}). If the substitution above is done in terms of the CMF function $\tilde{\Psi}$ by using the relationship (\ref{rwcmf}) instead of the Regge Wheeler function all components of (\ref{conodd}) are trivally satisfied. This is because the relationship, (\ref{rwcmf}), between the CMF function and the Regge Wheeler function assumes that the master equation is satisfied.

\subsection{Conservation of even stress-energy tensor}
As mentioned in the beginning of the previous section, we must keep the subtleties of the trace in mind when using (\ref{stresstensor}) to calculate the even perturbation of the stress-energy tensor. Since the expression is lengthy we have included it in the Appendix. 

We use the gauge condition $G=j_t=j_r=0$ and, in anaolgy with the odd case, we invoke the conservation equations (\ref{conodd}). Keeping only the linear terms we find that the $\nu = t$ component of (\ref{conodd}) gives,
\begin{align}\label{eventpart}
\bigg[& \left( \frac{\sqrt{f}}{r} - \frac{f'}{2 \sqrt{f}} \right) \partial_t K - \frac{\lambda \sqrt{f} }{2r^2} h_{tr} \\
& - \frac{f \sqrt f }{r}  \partial_t h_{rr} + \sqrt{f} \partial_r \partial_t K \bigg]Y =0 \nonumber
\end{align}
which is  the same as the $tr$ component of the Einstein equations. Using (\ref{eventpart}) and (\ref{htt})$-$(\ref{Z}), it can be shown that the $\nu=A$ components yield:
\begin{equation}
\left(\square  - V_{even}  \right)\tilde{\Phi}_{even} Y_A = 0
\end{equation}
which is equivalent to the even master equation, (\ref{mastereven}).

\section{Fluid representation}
In this section we show how the stress-energy tensor, along with its perturbation, can be expressed in a fluid form, determined entirely from the master function. This allows us to connect fluid properties like energy, velocity, viscosity, and vorticity  with gravitational quantities of the bulk. We will make this connection both at finite-$r$ surfaces and at infinity. 

To begin, we briefly review the fluid stress-energy tensor (we closely follow \citep{Bakas:2008gz}).  For perfect fluids we have
\begin{equation}\label{perf}
T_{\mu\nu} = \mathcal{E} u_{\mu} u_\nu + \mathcal{P} \Delta _{\mu\nu}
\end{equation}
where, $\enr$ is the energy density, $\prs$ is the pressure, and $\Delta_{\mu\nu}= u_{\mu} u_\nu + \gamma _{\mu\nu}$. As $r\rightarrow \infty$ the trace of this stress tensor vanishes however
at finite-$r$ this is not generally the case. Instead we have the following  equation of state:
\begin{equation}\label{eosstatic}
\prs=\frac{1}{2} \left(T+\enr \right)
\end{equation}
where $T$ is the trace of (\ref{perf}). To include the effects of dissipation the stress-energy tensor may be written as 
\begin{equation}
\bar{T}_{\mu\nu} = \bar{\enr} u_{\mu} u_\nu + \bar{\prs} \bar{\Delta} _{\mu\nu} + \Pi_{\mu\nu}.
\end{equation}
We have added a `$-$' over quantities to show that there may be linear perturbations to the metric, energy and pressure. Note that now the equation of state is also modified to include perturbations,
\begin{equation}\label{eos}
\bar{\prs} = \frac{1}{2}\left( \bar{T}+ \bar{\enr} \right)
\end{equation}
and we will be working in the Landau frame,
\begin{equation}\label{landau}
\bar{T}_{\mu\nu} u^\nu = -\bar{\enr} u_\mu.
\end{equation}
 The quantity $\Pi_{\mu\nu}$ is transverse to the velocity and captures the viscous effects of the fluids and can be expanded in terms of the derivatives of the velocity:
\begin{equation}
\Pi_{\mu\nu} = \Pi^{(1)}_{\mu\nu} + \Pi^{(2)}_{\mu\nu} + \hdots 
\end{equation}
where the superscripts denote the order of the derivative of $u_{\mu}$. We will only be interested in the first order,
\begin{equation}
\Pi_{\mu\nu}^{(1)} = -\eta \sigma_{\mu\nu} - \zeta \bar{\Delta}_{\mu\nu} \bar{\mathcal{D}}_\gamma u^ {\gamma} 
\end{equation}
with $\eta$ as the shear viscosity and $\zeta$ is the bulk viscosity which we will take to be zero. This leaves us with $-\eta \sigma_{\mu\nu}$ for
\begin{equation}
\sigma_{\mu\nu} = 2 \bar{\mathcal{D}}_{<\mu} u_{\nu>} 
\end{equation}
and
\begin{equation}
\bar{\mathcal{D}}_{<\mu} u_{\nu>} = \bar{\Delta}_{\mu\sigma} \bar{ \Delta}_{\nu\gamma} \bar{\mathcal{D}}^{(\sigma} u^{\gamma)} - \frac{1}{2} \bar{\mathcal{D}}_{\mu\nu} \bar{\Delta}_{\sigma \gamma} \bar{\mathcal{D}}^\sigma u^\gamma.
\end{equation}
Hence, $\sigma_{\mu\nu}$ is the transverse, symmetric, and traceless part of $\Pi_{\mu\nu}^{(1)}$. We will also make use of the following formula for the anti-symmetric vorticity tensor:
\begin{equation}
\bar{\omega}_{\mu\nu} = \bar{\Delta}_{\mu\sigma} \bar{ \Delta}_{\nu\gamma} \bar{\mathcal{D}}^{[\sigma} u^{\gamma]}
\end{equation}
It was shown in \citep{Bakas:2008gz} that the vorticity of the fluid vanishes at infinity when even perturbations were used. Below, we confirm that this result continues to hold on finite-$r$ surfaces. 

\subsection{Fluid representation of the static stress-energy tensor}
We can quite easily get the fluid representation for the static case by using the Landau condition (\ref{landau}) with (\ref{oddstess1}) and (\ref{oddstress2}) as the stress-energy tensor. By taking $\mu=t$ in (\ref{landau}) we find the energy density to be
\begin{equation}\label{staticenergy}
\enr = \tau_1 \, . 
\end{equation}
The trace is given by, $T=2\tau_2 - \tau_1$.
Thus, by using the equation of state (\ref{eosstatic}) we find the pressure to be
\begin{equation}\label{staticpressure}
\prs=\tau_2 \, . 
\end{equation}
Finally, by taking the $\mu=A$ in (\ref{landau}) we have
\begin{equation}\label{staticvelocity}
u_A =0
\end{equation}
and $u_t=-\sqrt{f}$ by requiring that $u^{\mu}u_{\mu}=-1$.

\subsection{Fluid representation of the  stress-energy tensor}
To find the fluid representation with odd perturbations we use the CMF function (\ref{CMF}). The form of the CMF function is particularly useful in simplifying the odd perturbations to the stress-energy tensor. In the gauge choice $h_2=0$, in terms of $\Psi$ we have
\begin{align}
\delta T_{tA}= & \left(A \Psi + B \partial_{r^*} \Psi\right)X_A \ \ \text{and}\\
\delta T_{AB}= & \partial_t \left(C \Psi + D \partial_{r^*} \Psi\right)X_{AB}
\end{align}
where $A,B,C$ and $D$ are functions of $r$ and are defined in the Appendix. Notice that we have removed the `$\sim$' symbol from $\Psi$ to emphasize that a gauge choice has been made and one can only use equations (\ref{psirwdef1}) and (\ref{psirwdef2}) to get the metric components $h_t$ and $h_r$, and not the gauge invariant quantities, $\tilde{h}_t$ and $\tilde{h}_r$. 
We can get the fluid representation of the odd stress-energy tensor by using the Landau frame (\ref{landau}) which allows us to find the energy density and the velocity of the fluid. The requirement for the fluid to be timelike gives
\begin{equation}
u_t=-\sqrt{f} \, .
\end{equation} 
The energy density is the same as the static case:
\begin{equation}
\enr= \tau_1
\end{equation}
and since the trace of the stress-energy tensor is the same as the static case, the pressure is the same as (\ref{staticpressure}).
The spatial components of the velocity are $u_A= U_{odd} X_A$, where:
\begin{equation}
U_{odd}=-\chi \left[ \left(A - \frac{f}{2} \tau_2 \right)\Psi + \left(B -\frac{r}{2} \tau_2 \right) \partial_{r^*} \Psi \right]
\end{equation}
with
\begin{equation}\label{chi}
\chi= \frac{1}{\sqrt{f}\left(\enr + \prs\right)} \, .
\end{equation}
The shear tensor of this velocity field is $\sigma_{AB}= \Sigma_{odd} X_{AB}$ where
\begin{equation}
\Sigma_{odd} =2U_{odd} -\sqrt{f} \Psi - \frac{r}{\sqrt{f}}\partial_{r^*} \Psi  \, . 
\end{equation}
Since there are no perturbations to the energy and pressure, and we take $h_2=0$, we get
\begin{equation}
\Pi^{(1)}_{AB} = \delta T_{AB}.
\end{equation} 
Further since both $\sigma_{AB}$ and $\delta T_{AB}$ are proportional to $X_{AB}$, we can find the viscosity
\begin{equation}
\eta_{odd} = -\Sigma^{-1}_{odd}\partial_t \left(C \Psi + D \partial_{r^*} \Psi\right).
\end{equation}
Finally, it was found in \citep{Bakas:2008gz} that the odd vorticity is non-vanishing at the boundary. We find that on any finite-$r$ surface the vorticity is
\begin{equation}
\omega_{AB} =U_{odd} \mathring{X}_{AB} 
\end{equation}
where $\mathring{X}_{AB}$ is an anti-symmetric tensor defined in the Appendix.

\subsection{Fluid representation of the even stress-energy tensor}
To find the fluid representation with even perturbations we proceed by using the gauge condition $G=j_t=j_r=0$ and use (\ref{htt})$-$(\ref{Z}) to write the stress-energy tensor in terms of $\Phi$:
\begin{equation}
\delta T_{tt} = \left(E_1 \Phi + E_2 \partial_{r^*} \Phi + E_3 \partial_t^2 \Phi\right)Y,
\end{equation}
\begin{equation}
\delta T_{tA} =  \partial_t (F_1 \partial_{r^*} \Phi + F_2  \Phi)Y_A \ \ \text{and}
\end{equation}
\begin{align}\label{evenAB}
\delta T_{AB} =& \left[ G_1 \Phi + G_2 \partial_{r^*} \Phi + \partial_t^2 (G_3 \partial_{r^*} \Phi + G_4 \Phi) \right] \Omega_{AB} Y \nonumber \\
&  + \left(G_5\Phi + G_6 \partial_{r^*} \Phi + G_7 \partial_t^2 \Phi \right) Y_{AB}
\end{align}
where the $E$'s, $F$'s and $G$'s are functions of $r$ and are defined in the appendix.
Note that we have again removed the `$\sim$' symbol, emphasizing that equations (\ref{htt})$-$(\ref{Z}) may only be used to find $h_{ab}$ and $K$, and not $\tilde{h}_{ab}$ and $\tilde{K}$.   Continuing like we did for the odd case, we use the Landau condition to find the energy density and velocity. Requiring the fluid velocity be timelike gives,
\begin{equation}
u_t = -\sqrt{f} + \frac{1}{2 \sqrt{f}} h_{tt} Y.
\end{equation}
The perturbation to the energy density $\delta \enr$ is given by:
\begin{equation}\label{perenr}
\delta \enr = \frac{1}{f}\left( E_1 + \frac{1}{2} f\tau_1 r V_{even} \right) \Phi + \frac{1}{f} \left( E_2 -\frac{\tau_1 Q_-}{8H} \right)\partial_{r^*} \Phi.
\end{equation}
The spatial components of the velocity are $u_A=U_{even}Y_{A}$, where
\begin{equation}
U_{even}=-\chi \partial_t \left(F_1 \partial_{r^*} \Phi + F_2  \Phi  \right).
\end{equation}
The shear tensor of  this velocity field is $\sigma_{AB}= \Sigma_{even} Y_{AB}$ where
\begin{equation}
\Sigma_{even} = 2U_{even}.
\end{equation}
To find the viscosity we recall that $\Pi^{(1)} _{AB}$ is a trace-free tensor on the sphere. So we expect that, $\Pi^{(1)} _{AB}= (\hdots) Y_{AB}$. To show this note that since the energy, trace, and hence the pressure have  perturbations, we have
\begin{equation}
\Pi^{(1)} _{AB} = \delta T_{AB} - \prs \delta \gamma _{AB}  - \delta \prs \gamma_{AB} 
\end{equation}
using the equation of state (\ref{eos}) and the perturbation to the energy (\ref{perenr}) it can be shown that
\begin{equation}\label{tracefree}
\Pi^{(1)} _{AB} = \delta T_{AB} - \frac{1}{2} \Omega_{AB} \delta T_{FG} \Omega^{FG}
\end{equation}
which is the trace-free part of (\ref{evenAB}). Now, given that the shear tensor is also proportional to $Y_{AB}$, we may use (\ref{tracefree}) to find the viscosity
\begin{equation}
\eta_{even} = -\Sigma^{-1} \left(G_5\Phi + G_6 \partial_{r^*} \Phi + G_7 \partial_t^2 \Phi \right).
\end{equation}
We also found that the vorticity of the even perturbations vanishes, in agreement with results at infinity of \citep{Bakas:2008gz}.

%Next we choose the Regge wheeler gauge, $h_{2}=0$, $\tilde{h}_{t}=h_{t}$
%and $\tilde{h}_{r}=h_{r}$, and use (\ref{psirwdef}) to express the stress-energy tensor in terms of $\Psi_{RW}$;
%\begin{align}
%\delta T_{tA}= & \frac{\sqrt{f}}{2}\left(\partial{}_{t}\tilde{h}{}_{r}-\partial_{r}f\int\left(r\Psi_{RW}\right)'dt'+\frac{2}{r}f\int\left(r\Psi_{RW}\right)'dt'\right)X_{A}+\frac{L\mu}{2r^{2}}f\int\left(r\Psi_{RW}\right)'dt'X_{A}+\tau_2f\int\left(r\Psi_{RW}\right)'dt'X_{A}\\
%\delta T_{AB}= & \left(\sqrt{f}\tilde{h}_{r}-\frac{L}{f}\partial_{t}\tilde{h}_{t}\right)X_{AB}+
%\tau_2h_{2}X_{AB}
%\end{align}
%Further we have the following from \cite{Martel:2005ir};
%\begin{align}
%h_{t}= \frac{1}{2}& f\left(r\Psi_{odd}\right)'\\
%h_{r}= \frac{1}{2}&\frac{r}{f}\partial_t\Psi_{odd}
%\end{align}
%Then we can write the components of the stress-energy tensor as,
%\begin{align}
%\delta T_{tA}= & \left[-\frac{\sqrt{f}}{4}\frac{\mu}{r}\Psi+\left(\tau_2+\frac{L\mu}{2r^{2}}\right)%\frac{f}{2}\left(r\Psi\right)'\right]X_{A}\\
%\delta T_{AB}= & \left[\frac{r}{2\sqrt{f}}\partial_t\Psi-\frac{L}{2}(r \partial_t \Psi)'\right]X_{AB}
%\end{align}

\subsection{Fluid Representation on boundary}

In this section we show how we can take the above fluid representation of the fluid on finite-$r$ surfaces to the surface where $r \rightarrow \infty$. Taking this limit we have the following normalization factors:
\begin{align*}
&\mathbf{\bar{T}}_{\mu\nu} = \lim_{r\rightarrow \infty} \frac{r}{L} \bar{T}_{\mu\nu},
 &\pmb{\bar{\gamma}}_{\mu\nu} =  \lim_{r\rightarrow \infty} \frac{L^2}{r^2} \bar{\gamma}_{\mu\nu},  \\
&\mathbf{u}_{\mu} = \lim_{r\rightarrow \infty} \frac{L}{r} u_{\mu},
 &\pmb{\eta} = \lim_{r \rightarrow \infty} \frac{r^2}{L^2} \eta \ \ \text{and} \\
&\pmb{\bar{\mathcal{E}}} =\lim_{r \rightarrow \infty} \frac{r^3}{L^3} \mathcal{E}.
\end{align*}
This allows us to write down formulas for the fluid quantities at the boundary in terms of the odd and and even master functions at infinity. The stress-energy tensor at the boundary for the odd case is:
\begin{align}
\delta\mathbf{ T}_{tA}&= \frac{1}{2L^2}\left( M \Psi_{\infty} +\frac{1}{2}\mu L^2 \partial_{r^*}\Psi_\infty \right)X_{A} \label{oddbound}\\
\delta\mathbf{T}_{AB}&= -\frac{1}{2}L^2\partial_{r^*}\dot{\Psi}_\infty X_{AB}
\end{align}
where $\Psi_\infty := \Psi(t,r=\infty)$, $\partial_{r^*}\Psi_\infty = \partial_{r^*} \Psi (t,r=\infty) $ and $\partial_{r^*}\dot{\Psi}_\infty := \partial_t\partial_{r^*}\Psi(t,r=\infty)$.
The components of the velocity for the odd case are
\begin{equation}
\mathbf{u}_t=-1 \;  \quad \text{and} \quad
\mathbf{u}_A=-\frac{\mu L^2}{12M}\partial_{r^*}\Psi_\infty X_A \,. 
\end{equation}
Finally the viscosity for the odd case is,
\begin{equation}
\pmb{\eta}_{odd}=-\frac{3ML^2 \partial_{r^*} \dot{\Psi}_\infty}{\mu  L^2\partial_{r^*} \Psi_\infty+6M\Psi_\infty}.
\end{equation}
Similarly for the even case we have the stress-energy tensor at infinity:
\begin{align}
\delta \mathbf{T}_{tt} =& \left[ \left(\frac{\mu\lambda}{4L^2}   + \frac{18 M^2}{L^4 \mu} \right) \Phi_{\infty} -\frac{3M}{L^2} \partial_{r^*} \Phi_{\infty} \right]Y \\
\delta \mathbf{T}_{tA} =& - \frac{\mu}{4}\dot{ \Phi}_{\infty} Y_A \\
\delta \mathbf{T}_{AB} =&\left[ \left(\frac{\mu\lambda}{8} + \frac{3M^2}{L^2 \mu} \right) \Phi_{\infty} - \frac{1}{2} M \partial_{r^*} \Phi_{\infty} \right ] \Omega_{AB} Y \\
 &+ \left[\left( \frac{\lambda}{4} + \frac{18M^2}{\mu^2L^2} \right) \Phi_\infty - \frac{3M}{\mu} \partial_{r^*} \Phi_{\infty} + \frac{L^2}{2} \ddot{\Phi}_\infty \right] Y_{AB}. \nonumber
\end{align}
The perturbation to the energy for the even case is
\begin{equation}
\delta \pmb{\mathcal{E}} = \left( \frac{\mu\lambda}{4L^2} + \frac{18 M^2}{L^4 \mu} \right) \Phi_\infty - \frac{3M}{L^2} \partial_{r^*} \Phi_\infty.
\end{equation}
The velocity for the even case is
\begin{equation}
\mathbf{u}_t=-1 \quad \text{and} \quad
\mathbf{u}_A=\frac{L^2 \mu}{12 M} \dot{\Phi}_{\infty}Y_A 
\end{equation}
and finally the viscosity for the even case is,
\begin{equation}\label{evenvisc}
\pmb{\eta}_{even}= -\frac{6M}{L^2 \mu \dot{\Phi}_\infty} \left[\left( \frac{\lambda}{4} + \frac{18M^2}{\mu^2L^2} \right) \Phi_\infty - \frac{3M}{\mu} \partial_{r^*} \Phi_{\infty} + \frac{L^2}{2} \ddot{\Phi}_\infty \right].
\end{equation}
When we go to the frequency domain we find that expressions (\ref{oddbound})$-$(\ref{evenvisc}) agree with those presented in \cite{Bakas:2008gz}.

%{\color{red} Maybe a paragraph here about large/small and whether these are really fluids}
%
%{\color{blue} Not sure what to say here, using the boundary conditions of \citep{Bakas:2008gz}, reflecting for odd and mixed for even we get,
%\begin{equation}
%\pmb{\eta}_{odd}=-\frac{3M}{\mu}\frac{\partial_{r^*} \dot{\Psi}_\infty}{  \partial_{r^*} \Psi_\infty}
%\end{equation}
%and for odd,
%\begin{equation}
%\pmb{\eta}_{even}= =-\frac{3M}{\mu} \left(\frac{C_0 \Phi_\infty+\ddot{\Phi}_\infty }{ \dot{\Phi}_\infty} \right)
%\end{equation}
%where $C_0=\frac{\lambda}{2L^2}$. Now from here if we plug in the frequencies,
%\begin{equation}
%\omega_{odd} = -i \frac{3}{\mu} \frac{1}{r_h}
%\end{equation}
%and
%\begin{equation}
%\omega_{even} =\pm \sqrt{C_0} -i \frac{\mu}{6} \frac{1}{r_h}
%\end{equation}
%for $r_h\rightarrow \infty$. In this limit we get the correct result,
%\begin{equation}
%\pmb{\eta}_{odd/even}= \frac{m}{r_h}.
%\end{equation}
%So basically as long as the coefficients of $-\frac{3M}{\mu}$ are constant and real we have a well-defined viscosity. But what about the energy and equation of state? Do they cooperate with the viscosity? Atleast in the frequency domain they do \citep{Bakas:2008gz}.
%}
\section{Conclusion}
As a result of a number of studies over the past few years, the fluid/gravity correspondence has come to be understood in a precise sense  in particular for the brane (non-compact flat horizon) case. 
The aim of the present work was to explore the emergence of the duality from the viewpoint of standard perturbation theory.  The dynamics of perturbations of spherical AdS black holes and their corresponding fluid 
interpretation were analyzed within the robust classical framework that describes perturbations of spherically symmetric black hole spacetimes.  A key feature of this formalism is that it is covariant 
on the orbits of spherical symmetry (i.e. in the $(t,r)$ coordinates). This allows one to avoid explicitly working in the frequency domain. 

From this  perspective an important question is: under what conditions do the gravitational perturbations have an equivalent description as a 
near-equilibrium fluid?  This is certainly possible for large ($M \gg L$) black holes if the fluid is taken to live at timelike infinity on $\mathbb{R} \times S^2$. However even on
surfaces of finite-$r$ some  fluid-like behaviours remain. For example, the conservation of the quasilocal BYBK stress-energy tensor follows from a geometric identity that holds on all surfaces 
irrespective of the size of the black hole. Thus on any such surface the stress-energy tensor can be viewed as arising from some kind of matter that obeys conservation laws. 
Further, one can always write the stress-energy tensor in a fluid-like form. The real question then is: under what circumstances does that interpretation make sense so that the stress-energy
evolves in the same way as that of a fluid?

In an effort to better understand the emergence of fluid behaviour we have reformulated as much of the problem as possible in terms of the well-developed perturbation theory of spherical spacetimes. 
We have seen that components of the stress-energy can be rewritten in terms of the master functions and the conservation laws are equivalent to the master equations.
Various fluid quantities can then also be written in terms of the master function.  In particular one can show that the expressions for viscosity match those found \cite{Bakas:2008gz} when one restricts to the frequency domain and sends $r\to \infty$. 

A natural further investigation would  perform a numerical integration for the master function along the lines of \cite{Wang:2000dt}. Doing this for a range of parameters and studying the
BYBK stress-energy tensor on surfaces of increasingly large $r$ would allow one to study  the emergence of ``fluidness''. More precisely one could determine the circumstances under
which the identifications from Section IV produce a genuinely physical fluid. 

 As an example of non-fluid behaviour, note that  even in the frequency domain one does not get a real viscosity for all QNMs. This emerges only for the case of low-lying modes \cite{Michalogiorgakis:2006jc}. Taking the odd viscosity as an example,
\begin{equation}
\pmb{\eta}_{odd}=-\frac{3ML^2 \partial_{r^*} \dot{\Psi}_\infty}{\mu  L^2\partial_{r^*} \Psi_\infty+6M\Psi_\infty}
\end{equation}
for the boundary condition $\Psi_{\infty} =0$ and assume $\Psi= R(r)e^{-i\omega t}$ then $\partial_{r^*} \Psi_{\infty} = K_1 e^{-i\omega t}$ and $\partial_{r^*} \dot{\Psi}_{\infty} =-i\omega K_1 e^{-i\omega t}$. Hence,  in the frequency domain 
\begin{equation}
\pmb{\eta}_{odd}=i\omega\frac{3M }{\mu  }.
\end{equation}
If $\omega$ has a real part (for large black holes this frequency is purely imaginary \citep{Cardoso:2001bb},\citep{Bakas:2008gz}) the viscosity is imaginary and so the identification of the BYBK
stress-energy as that of a fluid fails.

We have made an initial attempt to implement this numerical integration.  
Unfortunately while our code ran well for small black holes, it developed numerical difficulties precisely during the transition to large black holes (where
the required resolution at large $r$ became impossibly fine).  Similar problems were previously encountered in \cite{Morgan:2009pn}. 
We will return to those issues in the future, but for now settle for having established the  foundation from which those studies may proceed.

\section{Appendix}
\subsection{Odd and even stress-energy tensor}
The even stress-energy tensor in terms of the metric perturbations in a general gauge is given by:
\begin{align}
\delta T_{tt}=&\left( \frac{3\sqrt{f}}{r} -\frac{L}{r^2} -\frac{2}{L} \right)f^2 \tilde{h}_{rr} Y + \frac{\mu f L}{2r^2} \tilde{K}Y  \\
& -f \sqrt{f} \partial_r \tilde{K} Y -2\tau_1 \partial_t \varepsilon_t Y \nonumber \\ 
&+\left[ \left( \frac{L}{r^2} +\frac{2}{L} -\frac{3\sqrt{f}}{r} \right) f'  + \frac{\mu f L}{r^3} + \frac{2 f  \sqrt{f}}{r^2} - \frac{\sqrt{f} \lambda}{r^2} \right]f \varepsilon_r Y \nonumber \\
\delta T_{tA}=& \bigg[\frac{1}{2} \sqrt{f} \tilde{h}_{tr} -\frac{1}{2}L \partial_t \tilde{K}+\frac{1}{2} \tau_2 r^2 \partial_{t}G \\ & 
+\left( \sqrt{f}-\frac{Lf}{r}\right)\partial_{t}{\it \varepsilon_{r}}-\tau_1 {\it \varepsilon_{t}}\bigg]Y_A \nonumber \\
\delta T_{AB}=& \Bigg\{\left(\frac{1}{4}L f \lambda - \frac{3}{4} f' \sqrt f r^2 - \frac{1}{2} f^{3/2} r \right) \tilde{h}_{rr} \\
& - \frac{1}{2} f^{3/2} r^2 \partial_r \tilde{h}_{rr} + \frac{r^2}{\sqrt{f}}\partial_t \tilde{h}_{tr} -\frac{1}{2}\tau_2 r^2 \lambda G \nonumber 
\\ \nonumber &+ \tau_2 r^2 \tilde{K} + \left[ \left(\frac{1}{2} \sqrt{f} - \frac{1}{4} L f'\right) \lambda + f (r^2 \tau_2)'\right] \varepsilon_r  \nonumber \\
& + \left( \frac{r^2}{\sqrt{f}} - Lr \right) \partial_t^2 \varepsilon_r + \frac{1}{2} \sqrt{f} r^2 \partial_r\tilde{K} - \frac{1}{2} \frac{L r^2}{f} \partial_t^2 \tilde{K} \Bigg\} \Omega_{AB} Y \nonumber \\
& + \left[ \frac{1}{2} L f \tilde{h}_{rr} + \tau_2 r^2 G + \left(\sqrt{f} -\frac{1}{2} L f' \right) \varepsilon_r \right] Y_{AB} \nonumber
\end{align}
The radial functions defined for the odd stress-energy tensor are given by:
\begin{equation}
A=\frac{f}{2}\left( \frac{L\mu}{2r^2} + \tau_2 \right) -\frac{\sqrt{f} \mu}{4 r},  \quad \quad B=\frac{r}{2}\left( \frac{L\mu}{2r^2} + \tau_2 \right),
\end{equation}
\begin{equation}
C=\frac{r}{2\sqrt{f}} -\frac{L}{2} \quad \text{and} \quad D= -\frac{Lr}{2f}.
\end{equation}
The radial functions for the even stress-energy tensor are given by the functions:
\begin{align}
E_1 =& \left( \frac{\sqrt{f}}{2}-\frac{L}{2r}-\frac{r}{L} \right)V_{even} f \\ &+\left(\sqrt f + \frac{L\mu}{2r}  \right) \frac{f P_+ }{32H^2r^2} 
 -\frac{f\sqrt{f} P_+' }{32H^2r} + \frac{f\sqrt{f} P_+ H'}{16H^3r} \nonumber \\
E_2=&\frac{L\mu f}{2r^2} + \left(-\frac{3 \sqrt{f}}{r} + \frac{2}{L} +\frac{L}{r^2} \right)\frac{Q_-}{8H} - \frac{\sqrt{f} P_+}{32H^2r} \\
E_3=&-r\tau_1 
\end{align}
along with 
\begin{align}
F_1 =& \frac{1}{2} \left( \frac{r}{\sqrt{f}} -L \right) \\
F_2 =& -\frac{LP_+}{64H^2r} -\frac{P_Z}{8\sqrt{f}H}
\end{align}
and
\begin{align}
G_1=&\left( \frac{\sqrt{f}}{2} - \frac{2r}{L} + \frac{f'r}{2 \sqrt{f}}- \frac{\sqrt{f} H' r}{H}\right)\frac{P_+}{32H^2}  \\
& + \frac{Q_- V_{even}r^2}{16 \sqrt{f} H}+ \left(\frac{L\lambda r}{8}-\frac{f' r^3}{8\sqrt{f}}\right) V_{even} \nonumber \\ &- \frac{V_{even}'\sqrt{f} r^3}{4} + \frac{\sqrt{f}r}{2} \frac{P_+'}{32H^2} \nonumber \\
G_2=&\left( r \sqrt{f} - \frac{L \lambda}{2} -\frac{r^2 f'}{2\sqrt{f}} -\frac{\sqrt{f} H' r^2}{H}\right) \frac{Q_-}{16 f H} \\
& +\frac{rP_+}{64\sqrt{f}H^2} - \frac{ V_{even} r^3}{4 \sqrt{f}} + r^2 \tau_2 + \frac{Q_-'r^2}{16 \sqrt{f}H}  \nonumber \\
G_3 =& \frac{r^2}{2 f} \left( \frac{r}{\sqrt{f}} -L \right) \\
G_4 = & \frac{1}{f \sqrt{f}} \left( \frac{f' r^3}{4} + \frac{r^2 Q_-}{16 H} -\frac{P_Z r^2}{4  H} \right)   \\
 & + \frac{1}{f} \left( \frac{L\lambda r}{4} - \frac{LP_+ r}{64 H^2} \right)  - \frac{r^2}{2\sqrt f} \nonumber \\
G_5 =& \frac{rLV_{even}}{4} \\
G_6=& -\frac{LQ_-}{16fH} \\
G_7=& \frac{rL}{2f} 
\end{align}
where $P_+=P_X+P_Y$ and $Q_-=Q_X-Q_Y$.
\subsection{Even Harmonics}

This section closely follows \citep{Martel:2005ir}, we include it here for completeness. For the even scalar sector we have the usual spherical harmonic functions $Y^{lm}(\theta,\phi)$ which satisfy the equation $[\Omega^{AB} D_A D_B + l(l+1)] Y^{lm} = 0$. The even vector harmonics are defined as:
\begin{equation} 
Y_{A}^{lm} := D_A Y^{lm}, 
\end{equation} 
they satisfy the following orthogonality relations:
\begin{equation} 
\int \bar{Y}^A_{lm} Y^{l'm'}_A\, d\Omega =  
l(l+1)\, \delta_{ll'}\delta_{mm'}
\end{equation}
the bar indicates complex conjugation and $d\Omega :=\sin\theta\, d\theta d\phi$ is the area element on the unit sphere. The tensor harmonics are  $\Omega_{AB} Y^{lm}$ and
\begin{equation}
Y_{AB}^{lm} := \Bigl[ D_A D_B + \frac{1}{2} l(l+1) \Omega_{AB}
\Bigr] Y^{lm}
\end{equation}
they satisfy the following orthogonality relations:
\begin{equation} 
\int \bar{Y}^{AB}_{lm} Y^{l'm'}_{AB}\, d\Omega =  
\frac{1}{2} (l-1)l(l+1)(l+2)\, \delta_{ll'}\delta_{mm'}
\label{3.8}
\end{equation}
and are traceless:
\begin{equation} 
\Omega^{AB} Y_{AB}^{lm} = 0 
\end{equation}

\subsection{Odd Harmonics}
This section also closely follows \citep{Martel:2005ir}.
The odd scalar sector is empty since the scalar functions $Y(\theta,\phi)$ are even. The odd vector harmonics are defined as:
\begin{equation} 
X_{A}^{lm} := -\varepsilon_A^{\ B} D_B Y^{lm}. 
\label{3.2}
\end{equation} 
they satisfy the following orthogonality relations:
\begin{equation} 
\int \bar{X}^A_{lm} X^{l'm'}_A\, d\Omega =  
l(l+1)\, \delta_{ll'}\delta_{mm'},
\end{equation}
the bar indicates complex conjugation and $d\Omega :=\sin\theta\, d\theta d\phi$ is the area element on the unit sphere. The tensor harmonics are:
\begin{equation} 
X_{AB}^{lm} := -\frac{1}{2} \bigl( \varepsilon_A^{\ C} D_B +
\varepsilon_B^{\ C} D_A \bigr) D_C Y^{lm}. 
\label{3.7}
\end{equation} 
they satisfy the following orthogonality relations:
\begin{equation} 
\int \bar{X}^{AB}_{lm} X^{l'm'}_{AB}\, d\Omega =  
\frac{1}{2} (l-1)l(l+1)(l+2)\, \delta_{ll'}\delta_{mm'}.
\label{3.9}
\end{equation}
and are traceless:
\begin{equation} 
 \Omega^{AB} X_{AB}^{lm}=0
\label{3.11}
\end{equation}
We will also find it useful to define the following anti-symmetric tensor,
\begin{equation}
\mathring{X}_{AB}= D_{[A} X_{B]}
\end{equation}
The odd and even vector harmonics are orthogonal:
\begin{equation} 
\int \bar{Y}^{A}_{lm} X^{l'm'}_A\, d\Omega = 0,
\label{3.5}
\end{equation} 
as are the tensor harmonics:
\begin{equation} 
\int \bar{Y}^{AB}_{lm} X^{l'm'}_{AB}\, d\Omega = 0.
\label{3.10}
\end{equation} 

	\bibliography{fluid_gravity}

\end{document}